# How often does unguided peer interaction lead to correct response consensus? An example from Conceptual Survey of Electricity and Magnetism

Apekshya Ghimire and Chandralekha Singh

Department of Physics and Astronomy, University of Pittsburgh, Pittsburgh, Pennsylvania 15260, USA

**ABSTRACT**
In this research, we investigated the impact of peer collaboration and changes from individual to group performance of graduate students on the Conceptual Survey of Electricity and Magnetism (CSEM) without any guidance from the instructor. We define construction of knowledge as a case in which the group answered the question correctly but in the individual administration of the survey before the group work, one member gave the correct answer and the other gave incorrect answer. We find that there was a significant improvement in the performance of students after peer interaction, which was mostly attributed to construction of knowledge. However, students had very few opportunities to co-construct knowledge as there were hardly any situations in which neither student in a group provided a correct answer. We analyzed the effect size for improvement from individual to group scores for each CSEM item to understand the characteristics of these questions that led to productive group interaction. We also compared the group performance of the graduate students to the introductory physics students in a prior study using the CSEM to get insight into the concepts that showed differences for the two groups and those that were challenging for both groups of students before and after collaboration with peers. Our findings can motivate physics instructors to incorporate group interactions both inside and outside of the classroom even without instructor's involvement so that students at all levels can learn from each other and develop a functional understanding of the underlying concepts.

## 1. INTRODUCTION

Prior research suggests that encouraging and giving students opportunities to learn from their peers both inside and outside of the classroom may be a valuable strategy for aiding their learning, in addition to any scaffolding support offered by the course instructor.

Many educationists have advocated for the importance of peer collaboration without instructor guidance for over a century now. Dewey [1] developed a framework for participatory democracy that advocated for students to be provided with a supportive environment to work together and develop intellectually. He especially emphasized that these goals can be achieved by creating a space where students can work together as a team to construct knowledge in the absence of any authoritative figure. In addition, Hutchins described the importance of distributed cognition to complete an intellectual task so that cognitive resources are increased, and outcome is optimized [2, 3]. One-way distributed cognition can be accomplished is by giving students opportunities to work with peers as a group and overcome the limited working capacity by building upon each other's logic and organizing ideas [2-4].

Another important framework which emphasizes peer collaboration among pairs of students is the zone of proximal facilitation (ZPF) model [5]. This model predicts the success of collaborative learning when students have some knowledge in the area, but they cannot complete the task on their own [5, 6]. The framework posits that students can succeed at more complex problems that they cannot do individually by working with other students and taking advantage of collective expertise. If the task is beyond the competence of a group defined by the ZPF, students may not benefit from peer collaboration. But they can solve complex problems if the task is in the region of ZPF and in that case, adequate group competence can lead to successful outcomes [5, 6]. The ZPF framework builds on Vygotsky's Zone of Proximal Development (ZPD), which is a zone defined by what students can do alone vs. with the help of instructors who design instruction to account for and build on students' prior knowledge [7]. If pairs of students collaborating with each other are in ZPF, the task can be viewed to be such that it is in the ZPD of the group but out of the ZPD of each student individually [5].

Research on the importance of peer collaboration without instructor guidance in various courses shows its usefulness in improving the performance of students in different areas. There are various reasons why peer collaboration needs to be encouraged among students both at the graduate and undergraduate levels. It has been found that students retain the knowledge they learned through group work better than by working alone [5, 8, 9]. These results from the peer collaboration are consistent with the distributed cognition and zone of proximal facilitation models [10]. Students can benefit from collaborative work if some members cue the relevant prior knowledge and clarify doubts. In cases in which students did not solve the problem individually, they can use combined knowledge to come up with a correct solution. During peer discussion, students can benefit from the logic of others [11], and they are also likely to think about their own ideas more deeply while explaining them to others, which can provide greater level of clarity to their own ideas and help to co-construct knowledge. Thus, if the classroom time is limited, promotion of out-of-classroom collaboration can be useful for helping students develop a functional understanding of the underlying concepts. This can allow instructors to focus on concepts that might be challenging and out of the ZPF so that they can provide scaffolding support in those areas [12].

Additionally, peer collaboration can help in the growth of students in other aspects such as scientific communication skills and the ability to work effectively with others. Students are more likely to express their ideas and vocalize their doubts with their peers rather than their instructors. They may feel more comfortable questioning other's reasoning while working with their peers instead of an authoritative figure. This can increase their confidence and help them to gain practice in scientific communication as well as critical thinking [13-20]. While working with peers, they are less likely to be worried about being judged and are likely to explain their reasoning without being anxious which might take up cognitive resources that could otherwise be used to solve the problem [21-23]. It can also have a positive impact on motivational beliefs like self-efficacy [24-26], which is positively correlated with the performance of the students in courses related to science, technology, engineering, and mathematics (STEM) [27-30].

The impact of peer interaction has been found to be highly positive in college-level physics instruction [31-35]. One of the widely used methods for peer collaboration in physics is Mazur's peer instruction method in which lectures are integrated with peer interaction. This approach allows students to collaborate in the classroom and work on physics related concepts using multiple-choice questions [31, 36]. It is a formative assessment approach which helps to improve student learning. In this approach, students also need to be more engaged in the lectures as they know that they will be held accountable during discussions with their peers, and they will have to explain their reasoning regarding various physics concepts [31, 36]. It has been found that self-efficacy plays an important role when students collaborate and explain their views to each other, while cementing their knowledge of physics concepts [33]. Prior research involving three or more students [37-40], e.g., using context-rich problems, shows that it is beneficial to assign roles to students while collaborating with each other such as those of a group leader, timekeeper, skeptic etc. However, previous research also acknowledges the benefits of allowing students to choose their partners while working as a pair, since the familiarity with the peer can help with the learning process [41, 42].

Previous research in our group also shows that peer collaboration in physics is beneficial. Research using physics surveys such as Conceptual Survey of Electricity and Magnetism (CSEM) has shown that the group performance is better than the individual performance for introductory physics students [15, 43]. It was also found that students retain their knowledge when CSEM was administered individually after peer collaboration [15, 43]. Students benefitted significantly from the group work as compared to those who did not work in a group. It was found that the students co-constructed knowledge approximately 30% of the time when working with peers after lecture-based instruction without any guidance from instructors [15, 43]. The co-construction of knowledge, which refers to cases in which students individually did not have the correct solution to a problem but were able to construct the correct solution with peers, can happen due to various factors. In cases in which both students had different incorrect answers individually, they are likely to explain their reasoning and determine the correct answer by figuring out the flaws in their initial approach. Moreover, even if both group members had the same incorrect answer for a problem, they may realize that they were unsure about their approach and discuss their reasoning with each other to converge on the correct answer via co-construction of knowledge.

All these frameworks and previous research studies inspired us to study the impact of unguided peer collaboration on graduate student performance on the CSEM survey discussed here. The CSEM is a validated survey which was used to investigate the effectiveness of peer collaboration among first-year graduate physics students. Since our research involves groups of two (and rarely three) students, we allowed students to choose whom they wanted to collaborate with on the CSEM survey after each student answered the questions individually.

As noted, prior research shows peer collaboration benefited the introductory physics students on the CSEM survey [15, 43]. Since many students at all levels have similar alternative conceptions about concepts covered in the CSEM survey, discussion with peers can be beneficial for students at all levels to help them develop a functional understanding of physics concepts. Moreover, since the group of students in our research consists of first year graduate students, they may have a stronger sense of community than the introductory students as they share similar experiences while taking multiple courses together during the semester when they collaborated on the CSEM survey. This might also positively impact peer collaboration [44]. It is very valuable to investigate the effectiveness of working with peers not only on the CSEM survey but other validated conceptual surveys as well.[1]

## 2. RESEARCH QUESTIONS

To investigate the impact of working with peers on physics graduate student performance on the CSEM, we investigated the following research questions when first year physics graduate students worked in groups after working individually:

RQ1. Does unguided peer interaction help to improve students' performance on the CSEM?
RQ2. How often do students choose the correct answer as a group when one of them chooses the correct answer individually?
RQ3. How often do students choose the correct answer as a group when none of them chooses the correct answer individually?
RQ4. What are the effect sizes for the items investigating different concepts on the CSEM on which less than 85% of the students provided the correct responses individually?
RQ5. How does the group performance of the graduate students compare to the group performance of the introductory students on the CSEM?

## 3. METHODOLOGY

### 3.1. Participants

The participants of this research investigation were from a large public institution in the USA. All participants were graduate students in a mandatory graduate introduction to teaching course in the first semester of the first year in the physics department, which meets for two consecutive class periods (1 hour 50 minutes) each week. The students were familiar with all the topics covered on the CSEM survey through their prior undergraduate courses. This survey was administered to the students during the first few weeks of the semester. Most of these students were working as teaching assistants for the introductory physics courses within the physics department. This task was designed to help them reflect upon the importance of collaborative learning. Since the CSEM survey was administered during the first few weeks of first semester and the graduate students take the core graduate electricity and magnetism course for Ph.D. students during the second semester, their knowledge of all these topics reflects learning primarily from earlier undergraduate courses. Data for the survey were collected over two different years and averaged together. The performance of students was observed both individually and in groups before and after peer interaction.

First, the students ($N_I$ = 42) worked individually on the CSEM survey for a period of approximately 50 minutes. All students used paper scantrons to bubble their responses to each question and returned the answer sheets to the instructor once they completed the survey so that they did not have access to them when working with peers. Then, the students were asked to choose a peer and work on the same survey in groups in the same class period ($N_G = 20$) for the same amount of time. Students, who knew each other, were allowed to choose their partners. Most of the groups had two students working together but two of them had three students. All the group members appeared to be actively engaged in discussions, and they worked together without any guidance from the instructor. These groups were working in the same classroom so the class was noisy in the second half of the class once students started working with peers but the possibility for them to overhear adjacent groups' discussions cannot be completely eliminated. In addition, the students were not given any feedback after the first individual administration of the survey, so they were not sure if the initial responses they had already turned in, to the instructor were correct. To understand the reasoning behind student difficulties with the concepts covered by CSEM, a total of 47 students in an upper-level undergraduate electricity and magnetism course were asked to provide written explanations over a two-year period for why they selected certain choices on the 14 most challenging CSEM problems. Since the data presented from graduate students in this study is from those who were in their first semester of the first-year graduate program, they had only taken upper-level undergraduate electricity and magnetism courses. Thus, upper-level students and graduate students had very similar preparation in electricity and magnetism. In addition, six graduate students were interviewed individually using a think-aloud protocol in which they were asked to answer the same 14 of the most challenging CSEM questions while thinking aloud. They were not disturbed while they provided their reasoning except when they became quiet for a long time and only at the end, we asked them for clarification of the points they had not made clear. The graduate student interview data provided reasoning like the written reasoning provided by upper-level students. We did not encounter any difficulties during the in-class data collection process which can be summarized as follows (a flowchart summarizing the entire research including written explanations and individual interviews with students is shown in figure 1):

- Semester-long graduate TA professional development course for first-year graduate students that convenes once each week for 1 hour 50 minutes class period.
- Early in the semester, in two separate years, CSEM survey was administered to students individually in the first 50 minutes and then in group in the second half of the same class period.
- Students were asked to engage in collaborative work to help them understand the benefits of collaborative learning.
- 42 students answered the CSEM questions individually in the first 50 minutes using paper scantrons.
- Students were asked to get together with a peer of their choosing (out of 20 groups, there were two groups of three) and work collaboratively on the CSEM questions without instructor support.
- Students engaged with their peers to solve the CSEM questions collaboratively on another paper scantron with the names of all students collaborating with each other listed on it (students did not have access to their scantrons with individual responses).
- Since groups worked in the same classroom, the class was noisy in the second half of the class once students started working with peers but the possibility for them to overhear adjacent groups' discussions cannot be completely eliminated.

---

[1] For example, in another work, we investigated the extent to which unguided peer interaction is effective in the context of Magnetism Conceptual Survey (MCS). The CSEM survey that students worked on with peers in the investigation discussed here is very different from the MCS. The findings using these different validated surveys elucidate student difficulties in those different contexts when working individually and the extent to which peer interaction in those different surveys (e.g., CSEM or MCS) can help them.

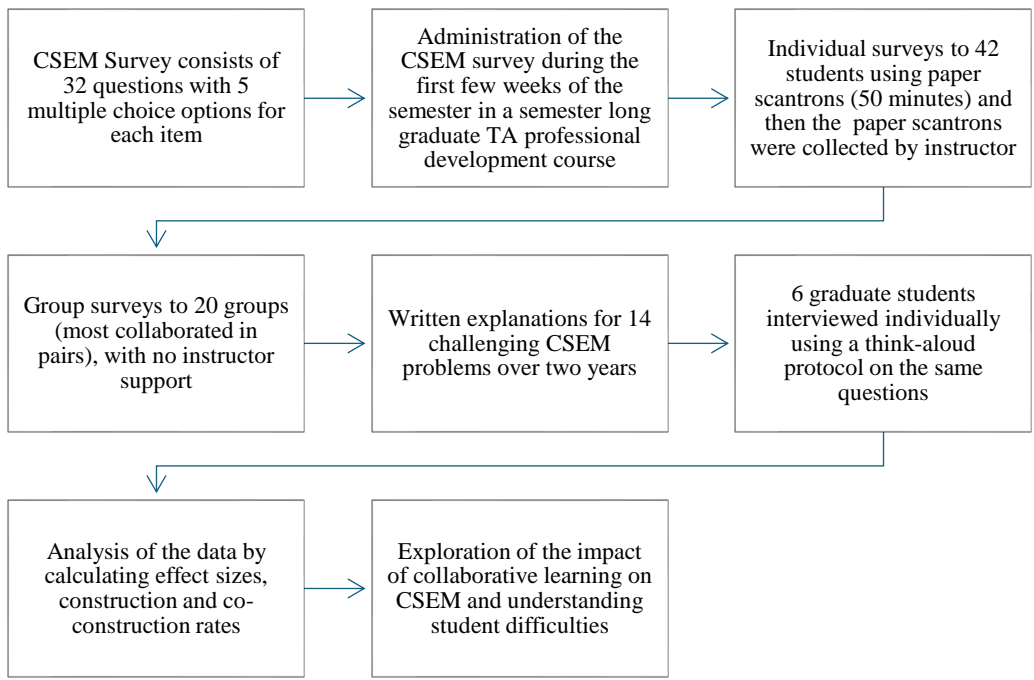

**Figure 1.** A Flowchart of the Research Process

### 3.2. Survey

The Conceptual Survey of Electricity and Magnetism (CSEM) is a validated survey which covers the topics, e.g., related to charge distribution on conductors/insulators, Coulomb's law, electric force, and electric field including superposition principle, electric potential and work, magnetic force, magnetic field and superposition principle, Faraday's law, Lenz's law and Newton's third law. The final version of this survey consists of 32 questions with five multiple-choice options for each item. It may be useful for the reader to have a copy of the survey even though we have tried our best to clarify the context of the questions.

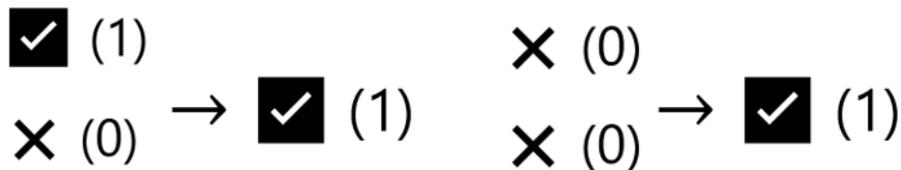

**Figure 2.** Diagrammatic Representation of Construction (left) and Co-Construction (right)

### 3.3. Analysis

The analysis of data was performed by calculating the percentage of students who selected the correct answers individually and in groups. The construction of knowledge for a CSEM item refers to a case in which the group answered the question correctly but in the individual administration of the CSEM before the group work, one member gave the correct answer and the other gave the incorrect answer (see figure 2). The rate of construction was also calculated for each question.

The rate of construction for each item is defined as the percentage of groups that gave the correct answer when the group consisted of a member who answered correctly and a member who answered incorrectly. The binary scores (0 for incorrect and 1 for correct) were used to write the combined scores for individuals and groups. In the notation we use, the individual score for a question is followed by the group score. For example, the first two digits in 01,1 or 10,1 refer to individual score (incorrect or correct) and the last score refers to the group score being 1. Thus, the formula used to calculate the rate of construction for each item is as follows in which N (10,1) refers to the number of groups that had the first student with correct response and the second student with incorrect response, but the group response was correct:

$$\frac{(N(10,1) + N(01,1))}{N(10,0) + N(10,1) + N(01,0) + N(01,1)} \times 100\%$$

Similarly, the rate of co-construction for each item is defined as the percentage of groups that gave the correct answer when none of the members gave the correct answer individually (see figure 2). In this case, the groups are assigned the binary scores of 00,1 where 00 represents the incorrect answers for individuals and 1 represents the correct answer for a group. The formula for calculating the rate of co-construction for an item is given below:

$$\frac{N(00,1)}{N(00,0) + N(00,1)} \times 100\%$$

In using the notation above, for the case of a few groups with three members, one of the repeated answer choices can be removed for individual choices before using these formulae.

Along with the rates of construction and co-construction, the effect sizes were also determined for each question to show the level of improvement[45]. In this study, the effect size is used to compare the individual and group scores for each item on the CSEM survey and investigate the level of improvement, if any, in the performance of the groups as compared to the individuals. The effect size, measured by Cohen's $d$, was divided into three categories[45]. Cohen's $d = (\bar{X}_1 - \bar{X}_2)/S_{pooled}$, where $\bar{X}_1$ and $\bar{X}_2$ refer to the sample means of individual score and group score and $S_{pooled}$ is the pooled standard deviation[45]. An effect was considered small if it is below 0.3, medium if it is between 0.3 and 0.6 and large if it is greater than 0.6.

## 4. RESULTS AND DISCUSSION

### RQ1. Does unguided peer interaction help to improve students' performance on the CSEM?

The answer to this research question is affirmative. Out of the 32 items of the CSEM survey, 19 items had an individual score less than 85%, which decreased to 6 items after peer interaction. As a measure of improvement, along with the standard errors, the average individual score was $(77 \pm 3)\%$ and the average group score was $(91 \pm 2)\%$. The individual and groups scores along with the standard errors for all 32 questions are shown in figures 3 and 4.

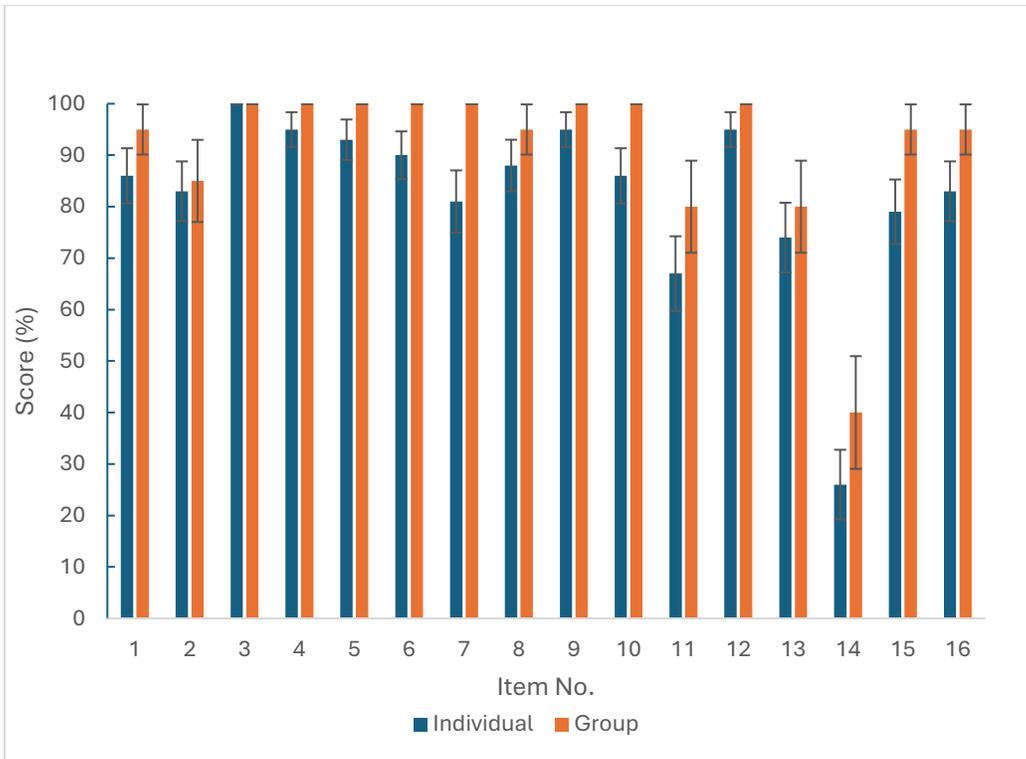

**Figure 3.** Individual and group scores for items 1-16 with standard errors

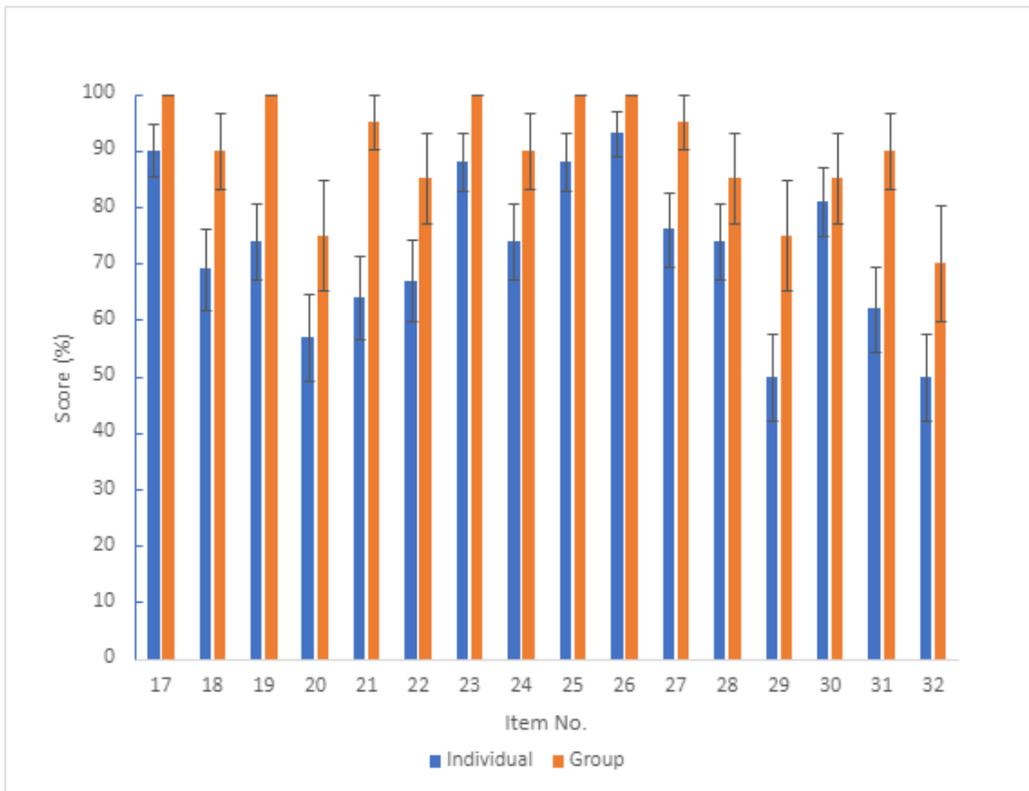

**Figure 4.** Individual and group scores for items 17-32 with standard errors

Table 1 shows the individual and groups scores, construction and co-construction rates and effect sizes along with the confidence intervals for each CSEM question. Also, Table 2 shows the distribution of students selecting each of the five answer choices individually and in groups in percentage for each question. Tables 1 and 2 show that there was a large variation in the individual score across all the CSEM questions. The lowest individual score was 26% for Q14 and the highest individual score was 100% for Q3. We adopt a heuristic that a score of 85% or greater corresponds to a good performance. Tables 1 and 2 show that there were three items with an individual score less than or equal to 50%. Only one item was between 50-60%, five items were between 60-70%, 6 items were between 70-80%, 9 items were between 80-90% and 8 items were between 90-100%. Tables 1 and 2 also show that most of these items had a group score greater than or equal to 85% except items Q11, 13, 14, 20, 29 and 32. The lowest group score was 40% for Q14 which was the most challenging question on the entire CSEM survey. These results are encouraging considering students were not given any feedback after the individual administration of the survey, and they did not get any feedback from the instructor during their group work either.

**Table 1.** The percentage of individuals and groups that chose the correct answer along with the rates of construction and co-construction and effect size along with the confidence interval (minimum and maximum values) for each item starting with item with highest effect size. The standard error for the construction rates ranged from 0% to 11% across all questions. All the percentages have been rounded up to the nearest whole number. '-' refers to no opportunity for co-construction i.e., there were no groups in which all students gave an incorrect answer individually.

| Item | Individual | Group | Construct (%) | Co-Construct (%) | Effect size | Min value | Max value |
| --- | --- | --- | --- | --- | --- | --- | --- |
| 21 | 64 | 95 | 100 | 67 | 0.74 | 0.16 | 1.32 |
| 19 | 74 | 100 | 100 | - | 0.70 | 0.12 | 1.28 |
| 31 | 62 | 90 | 90 | 67 | 0.64 | 0.08 | 1.2 |
| 7 | 81 | 100 | 100 | - | 0.58 | 0.02 | 1.14 |
| 29 | 50 | 75 | 100 | 0 | 0.51 | -0.05 | 1.07 |
| 18 | 69 | 90 | 100 | 0 | 0.50 | -0.06 | 1.06 |
| 27 | 76 | 95 | 90 | - | 0.50 | -0.06 | 1.06 |
| 10 | 86 | 100 | 100 | - | 0.48 | -0.08 | 1.04 |
| 23 | 88 | 100 | 100 | - | 0.44 | -0.12 | 1 |
| 25 | 88 | 100 | 100 | - | 0.44 | -0.12 | 1 |
| 15 | 79 | 95 | 89 | - | 0.43 | -0.13 | 0.99 |
| 32 | 50 | 70 | 91 | 0 | 0.41 | -0.15 | 0.97 |
| 6 | 90 | 100 | 100 | - | 0.40 | -0.16 | 0.96 |
| 17 | 90 | 100 | 100 | - | 0.40 | -0.16 | 0.96 |
| 22 | 67 | 85 | 83 | 0 | 0.40 | -0.16 | 0.96 |
| 24 | 74 | 90 | 89 | 0 | 0.39 | -0.17 | 0.95 |
| 20 | 57 | 75 | 70 | 50 | 0.38 | -0.18 | 0.94 |
| 16 | 83 | 95 | 86 | - | 0.35 | -0.21 | 0.91 |
| 5 | 93 | 100 | 100 | - | 0.32 | -0.22 | 0.86 |
| 26 | 93 | 100 | 100 | - | 0.32 | -0.22 | 0.86 |
| 14 | 26 | 40 | 70 | 10 | 0.30 | -0.24 | 0.84 |
| 1 | 86 | 95 | 100 | 0 | 0.28 | -0.26 | 0.82 |
| 4 | 95 | 100 | 100 | - | 0.28 | -0.26 | 0.82 |
| 9 | 95 | 100 | 100 | - | 0.28 | -0.26 | 0.82 |
| 11 | 67 | 80 | 75 | 33 | 0.28 | -0.26 | 0.82 |
| 12 | 95 | 100 | 100 | - | 0.28 | -0.26 | 0.82 |
| 28 | 74 | 85 | 86 | 0 | 0.26 | -0.28 | 0.8 |
| 8 | 88 | 95 | 80 | - | 0.23 | -0.31 | 0.77 |
| 13 | 74 | 80 | 67 | 0 | 0.14 | -0.4 | 0.68 |
| 30 | 81 | 85 | 67 | 0 | 0.10 | -0.44 | 0.64 |
| 2 | 83 | 85 | 60 | 0 | 0.05 | -0.49 | 0.59 |
| 3 | 100 | 100 | - | - | - | - | - |

Since student individual and group performances on each item of the survey give insight into what students were able to answer correctly individually or in group in each context, it is important to analyze these items individually to understand the extent to which the group work was effective. We do this type of analysis later in response to RQ4.

**RQ2. How often do students choose the correct answer as a group when one of them chooses the correct answer individually?**

We find that in 88% of the eligible cases across all questions, construction happened. Tables 1 and 2 provide details of construction for each question.

**RQ3. How often do students choose the correct answer as a group when none of them choose the correct answer individually?**

We find that in 19% of eligible cases across all questions, co-construction happened. Tables 1 and 2 provide details of co-construction for each question. We note that 15 out of 32 questions had an opportunity for co-construction since in other cases, the groups had at least one student in each group who had the correct answer individually before working in a group.

**RQ4. What are the effect sizes for the items investigating different concepts on the CSEM on which less than 85% of students provided the correct responses individually?**

The extent of improvement in the score of the students from before to after the peer interaction can be measured by using the effect size given by Cohen's *d* [45]. Q3, which is related to the determination of forces between two positive charges, was the easiest question on the CSEM survey on which all students provided the correct response individually. We divided all other questions into three categories based on their effect sizes and discuss them starting from the highest effect size questions as follows:

**High Effect Size**

Only three questions on the CSEM survey had a high effect size ($d > 0.6$) - Q19, 21 and 31. It is interesting to note that all these items had an individual score of less than 75% but the group score ranged from 90% to 100%.

Q21 is related to the motion of a charged particle placed at rest in a uniform magnetic field. This item had an effect size of 0.74 which is the highest among all CSEM questions. This item had a low individual score of 64% but it increased to a group score of 95%. There was 100% construction rate and 67% co-construction rate for Q21. Three groups had an opportunity for co-construction but only two groups converged to the correct answer. Individually, 19% of the students thought that the charged particle will move in a circle at a constant speed since the force is perpendicular to the velocity. Moreover, 10% of the students thought that it moves with a constant non-zero acceleration as the force is constant. In both cases, the students did not account for the fact that the charged particle is at rest, and a charge needs to be in motion to experience a force due to the magnetic field. One of the upper-level undergraduate students explained, "In a uniform magnetic field, charges will move in a circle of constant speed and radius proportional to a function of field and charge".

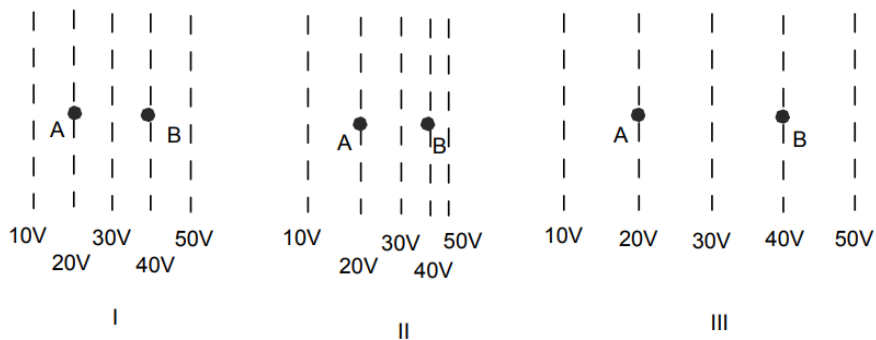

**Figure 5.** Figure provided for items 17-19

Q19 focuses on the direction of force exerted by the electric field on a positively charged particle when placed at two different points on equipotential surfaces shown in figure 5. It is the item with the second highest effect size of 0.70, which is due to 100% construction rate but there was no opportunity for co-construction as they were no groups in which both students gave an incorrect answer individually. The individual score for this item was 74% but the group work was so effective that all groups provided the correct answer. We find that 17% of students chose the direction opposite to the correct answer. Some students explained their reasoning using the concept of the motion of the charge. For example, one upper-level student explained, "The charge flows into the +Q and it will be in the direction of the motion of the point charge, which is to the right." However, all student groups converged on the correct answer after working with peers (see Tables 2 and 3).

In Q31, students are presented with a diagram of a neutral metal bar which moves with a constant velocity towards the right in a uniform magnetic field pointing out of the page and they must figure out the charge distribution on the metal bar. It is one of the items related to the force experienced by the charges in a metal bar moving in a uniform magnetic field. This item had an individual score of 62% and a group score of 90% with an effect size of 0.64. This item had a construction rate of 90% and the co-construction rate was 67%. Individually, 17% of the students chose the opposite direction to the actual direction predicted by using the right-hand rule incorrectly. Moreover, individually, 12% of students thought that there would be no charge distribution on the metal bar as the magnetic field is uniform. This suggests that some students did not know how to use the right-hand rule correctly to answer this question individually, but most groups converged on the correct response.

**Medium Effect Size**

We find that group performance improved compared to individual performance on 17 questions such that they had a medium effect size ($0.3 \leq d \leq 0.6$): Q6, 7, 10, 14, 15, 16, 17, 18, 20, 22, 23, 24, 25, 26, 27, 29, 32. Below, we will only present analysis of the items with individual scores below 85% to understand the characteristics of these questions, i.e., Q7, 14, 15, 16, 18, 20, 22, 24, 27, 29 and 32.

Q7 is related to the force between two charges (Newton's third law) which appears to be an easier concept for students as compared to others and within all group's ZPF since the group score was 100%. This item had an effect size of 0.58 as the performance increased from an individual score of 81% to a group score of 100%. The construction rate was 100% but there was no opportunity for co-construction, i.e., there were no groups in which both students gave an incorrect answer individually. Even though the score was less than 85% individually, all groups converged on the correct response. Table 2 shows that 14% of the students initially chose option C thinking that the electric force due to one charge is greater than the other. But after peer interaction, groups converged on the correct response that the forces acting on both charges are equal according to Newton's third law.

**Table 2.** The percentage of individuals and groups that chose each answer option for each item. The bold represents the correct answer for each question. Students who skipped a question (shown in column S) were marked as incorrect in that question for calculations of rate of construction and co-construction. (Number of individuals = 42, Number of groups = 20); "-" refers to no opportunity for co-construction i.e., cases in which there were no groups in which all students gave an incorrect answer individually.

| Item # | Group/Individual | A | B | C | D | E | S | Construction | Co-Construction |
|---|---|---|---|---|---|---|---|---|---|
| 1 | Individual | 2 | 86 | 12 | 0 | 0 | 0 | 100 | 0 |
|   | Group | 0 | 95 | 5 | 0 | 0 | 0 | | |
| 2 | Individual | 83 | 2 | 0 | 5 | 7 | 2 | 60 | 0 |
|   | Group | 85 | 0 | 0 | 15 | 0 | 0 | | |
| 3 | Individual | 0 | 100 | 0 | 0 | 0 | 0 | - | - |
|   | Group | 0 | 100 | 0 | 0 | 0 | 0 | | |
| 4 | Individual | 0 | 95 | 5 | 0 | 0 | 0 | 100 | - |
|   | Group | 0 | 100 | 0 | 0 | 0 | 0 | | |
| 5 | Individual | 7 | 0 | 93 | 0 | 0 | 0 | 100 | - |
|   | Group      | 0   | 0   | 100 | 0   | 0   | 0 |     |    |
| 6 | Individual | 5   | 2   | 2   | 0   | 90  | 0 | 100 | -  |
|   | Group      | 0   | 0   | 0   | 0   | 100 | 0 |     |    |
| 7 | Individual | 2   | 81  | 14  | 0   | 2   | 0 | 100 | -  |
|   | Group      | 0   | 100 | 0   | 0   | 0   | 0 |     |    |
| 8 | Individual | 2   | 88  | 2   | 2   | 2   | 2 | 80  | -  |
|   | Group      | 0   | 95  | 0   | 0   | 5   | 0 |     |    |
| 9 | Individual | 0   | 95  | 0   | 2   | 0   | 2 | 100 | -  |
|   | Group      | 0   | 100 | 0   | 0   | 0   | 0 |     |    |
| 10 | Individual | 2  | 5   | 86  | 5   | 2   | 0 | 100 | -  |
|   | Group      | 0   | 0   | 100 | 0   | 0   | 0 |     |    |
| 11 | Individual | 17 | 2   | 7   | 2   | 67  | 5 | 75  | 34 |
|   | Group      | 10  | 0   | 10  | 0   | 80  | 0 |     |    |
| 12 | Individual | 2  | 0   | 2   | 95  | 0   | 0 | 100 | -  |
|   | Group      | 0   | 0   | 0   | 100 | 0   | 0 |     |    |
| 13 | Individual | 17 | 10  | 0   | 0   | 74  | 0 | 67  | 0  |
|   | Group      | 15  | 5   | 0   | 0   | 80  | 0 |     |    |
| 14 | Individual | 48 | 12  | 0   | 26  | 12  | 2 | 70  | 10 |
|   | Group      | 50  | 5   | 0   | 40  | 5   | 0 |     |    |
| 15 | Individual | 79 | 10  | 7   | 0   | 5   | 0 | 89  | -  |
|   | Group      | 95  | 0   | 5   | 0   | 0   | 0 |     |    |
| 16 | Individual | 5  | 7   | 2   | 2   | 83  | 0 | 86  | -  |
|   | Group      | 0   | 0   | 0   | 0   | 95  | 5 |     |    |
| 17 | Individual | 0  | 5   | 5   | 0   | 90  | 0 | 100 | -  |
|   | Group      | 0   | 0   | 0   | 0   | 100 | 0 |     |    |
| 18 | Individual | 0  | 2   | 2   | 69  | 26  | 0 | 100 | 0  |
|   | Group      | 0   | 0   | 0   | 90  | 10  | 0 |     |    |
| 19 | Individual | 74 | 17  | 0   | 0   | 7   | 2 | 100 | -  |
|   | Group      | 100 | 0   | 0   | 0   | 0   | 0 |     |    |
| 20 | Individual | 2  | 14  | 19  | 57  | 7   | 0 | 70  | 50 |
|   | Group      | 0   | 5   | 20  | 75  | 0   | 0 |     |    |
| 21 | Individual | 5  | 10  | 19  | 2   | 64  | 0 | 100 | 67 |
|   | Group      | 0   | 5   | 0   | 0   | 95  | 0 |     |    |
| 22 | Individual | 2  | 12  | 19  | 67  | 0   | 0 | 84  | 0  |
|   | Group      | 5   | 0   | 10  | 85  | 0   | 0 |     |    |
| 23 | Individual | 88 | 2   | 2   | 5   | 2   | 0 | 100 | -  |
|   | Group      | 100 | 0   | 0   | 0   | 0   | 0 |     |    |
| 24 | Individual | 0  | 5   | 74  | 21  | 0   | 0 | 89  | 0  |
|   | Group      | 0   | 0   | 90  | 10  | 0   | 0 |     |    |
| 25 | Individual | 5 | 0 | 2 | 88 | 5 | 0 | 100 | - |
| | Group | 0 | 0 | 0 | 100 | 0 | 0 | | |
| 26 | Individual | 93 | 2 | 0 | 0 | 5 | 0 | 100 | - |
| | Group | 100 | 0 | 0 | 0 | 0 | 0 | | |
| 27 | Individual | 10 | 2 | 7 | 2 | 76 | 2 | 90 | - |
| | Group | 0 | 0 | 0 | 5 | 95 | 0 | | |
| 28 | Individual | 0 | 0 | 74 | 0 | 26 | 0 | 86 | 0 |
| | Group | 0 | 0 | 85 | 0 | 15 | 0 | | |
| 29 | Individual | 2 | 33 | 50 | 10 | 5 | 0 | 100 | 0 |
| | Group | 0 | 20 | 75 | 5 | 0 | 0 | | |
| 30 | Individual | 81 | 0 | 2 | 12 | 5 | 0 | 67 | 0 |
| | Group | 85 | 0 | 0 | 5 | 10 | 0 | | |
| 31 | Individual | 12 | 2 | 5 | 17 | 62 | 2 | 90 | 67 |
| | Group | 0 | 0 | 0 | 10 | 90 | 0 | | |
| 32 | Individual | 26 | 2 | 14 | 50 | 2 | 5 | 91 | 0 |
| | Group | 10 | 5 | 15 | 70 | 0 | 0 | | |

Q29 and Q32 are related to the use of Faraday's and Lenz's law. Both items were the most challenging for students after Q14 and they each had a low individual score of 50%. On Q29, students must identify the situations in which a current is induced in the loop as shown in figure 6. This item had an effect size of 0.51 with 100% construction rate and 0% co-construction rate. For two conditions (I and IV), there is a relative motion between the magnet and the current loop. And there is one condition (II) in which the current loop is collapsing. Table 2 shows that only 50% of the students considered all three conditions for induced current in the loop and 33% chose option B (I, IV) and 10% chose option D (IV). It appears that most of the students knew that the relative motion between the magnet and loop induces an emf (I and IV), but some had difficulty realizing that the change in the area of the loop (II) also leads to an induced emf. Table 2 also shows that 75% correctly chose option C (I, II, IV) after peer discussion but there were a few students who struggled with this concept even after working in groups. There were five groups with both individuals providing incorrect answers initially and none of them could converge on the correct solution as a group. In one group, two members initially chose option D, and after discussion, they stuck with option D, which is incorrect. In the remaining four groups, at least one student initially chose incorrect option B and all these groups converged incorrectly to option B after discussion with their peers. Thus, 20% of the groups chose option B after peer interaction. One upper-level student who chose B explained, "If either the magnet is moving or the copper wire is moving it will create a change in the current in the copper wire, causing the bulb to light, so in I and IV it will light, but not in II or III because they do not change locations in the magnetic field that is induced on the wire".

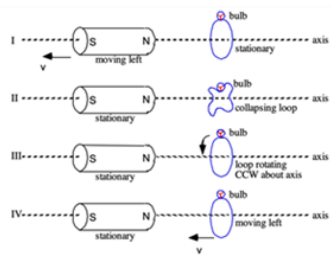

**Figure 6.** Figure provided with CSEM item 29

Q32 focuses on investigating student understanding of mutual induction and consists of two coils in which one is connected to an ammeter and the ammeter reading is given as shown in figure 7. The other coil is connected to a voltmeter and the students must find the correct time dependence of voltmeter reading. Since an emf is induced in the second coil due to change in the magnetic flux through the loop, the voltmeter reading would be positive and constant, zero, negative and constant, and zero again. Only 50% of the students provided the correct answer, which increased to a group score of 70% with an effect size of 0.40. This item had a construction rate of 91% and a co-construction rate of 0%. Table 2 shows that the most dominant incorrect choice was option A (28%) in which the voltmeter reading was the same as the ammeter reading followed by option C (15%) in which the voltmeter reading was reversed to the graph for the current. The correct answer, option D, shows that the voltmeter reading has a constant positive value when the ammeter reading increases, remains at zero when the ammeter reading is constant, maintains a constant negative value when the ammeter reading decreases, and returns to zero when the ammeter reading becomes constant again. There were five groups who had an opportunity for co-construction and none of these groups could converge to a correct answer after peer interaction. In the first group, one student originally selected option B while another chose option A, but they both ended up choosing incorrect option B after discussing it. The second group consisted of two students who individually selected option A and chose the same option after their discussion. Among the remaining three groups, each had one student who initially went for option C, and another who favored option A. In two of these groups, the students eventually converged on incorrect option C, while the third group converged on incorrect option A after discussing with their peers. Some students choosing A made an explicit connection between voltage and current using Ohm's law (voltage is proportional to the current when resistance is constant). For example, one upper-level student explained, "With a constant resistance the change in current will be reflected by the change in voltage through $V = I R$." Another student who chose C explained, "The ammeter shows the movement of charge, so the voltage should be negative as it is mirrored in the second coil."

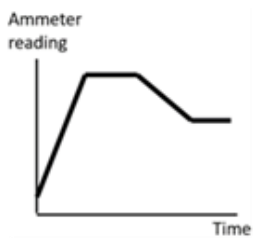

**Figure 7.** Figure provided with CSEM item 32

Q18 is related to the comparison of the electric field between two equipotential surfaces with same potential difference but with different distances between the surfaces as shown in figure 5. This item had a relatively low individual score of 69% and based upon the improvement from the individual to group performance, it had an effect size of 0.50. After group discussion, 90% of the groups provided the correct answer. This item had a 100% construction rate but 0% co-construction rate. Table 2 shows that 26% of the students chose option E thinking that all the three cases will have the same magnitude of electric field at point B when answering individually. Table 3 shows that there were two groups who were unable to converge on the correct solution with both students in the group with individual incorrect answers. These were the same students who chose option E individually and converged on the same option again after discussion with their peers. It appears that some students only considered the potential difference and neglected the distance between equipotential surfaces in their reasoning.

Q27 investigates whether a positively charged particle at rest placed between two bar magnets with opposite poles facing each other experiences a magnetic force. This item had an individual score of 76% and a high group score of 95% with an effect size of 0.50. It had 90% construction rate with no opportunity for co-construction. Table 2 shows that individually, 10% chose option A (right) and 7% chose option C (left). These students did not consider that the charge was not moving and chose one of the options for the directions of force. So, it appears that some graduate students are also struggling with the concept of magnetic force on charged particles. One upper-level student who chose option A explained, "The positive charge should be directed towards the south pole as that would be the net direction of the field" and another student who chose C explained, "one magnet pushes up and left, the other pulls down and left, for a net effect of left".

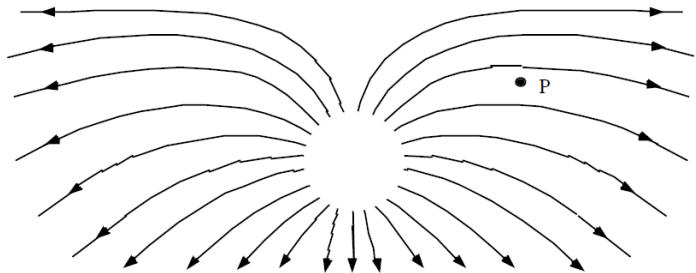

**Figure 8.** Figure provided with CSEM item 15

Q15 is related to the direction of the electric force on a negative charge in an electric field at point P shown in figure 8. This item had an individual score of 79% and it increased to a group score of 95%. The effect size was 0.43 with a high construction rate of 89% and there was no opportunity for co-construction. Table 2 shows that the dominant incorrect answers were B (southwest, 10%) and C (east, 7%). The students who chose option C did not recognize that negative charges experience a force in the opposite direction to the electric field. However, upon discussion with their partners, most groups converged on the correct direction for the electric force.

Q22 is related to the determination of direction of the magnetic field given the direction of velocity and motion of an electron. This item had an individual score of 67% and a group score of 85%. It had an effect size of 0.40 with 83% construction rate and 0% co-construction rate. There was only one group who had an opportunity for co-construction. Table 2 shows that options B and C were the most common incorrect answers (12% and 19%, respectively) chosen by students. Those who chose option B appear to be struggling with the direction of force, velocity, and magnetic fields in the right-hand rule whereas those who chose option C used the right-hand rule incorrectly. One upper-level student who chose option C explained, "Because the electron is negative, a B field pointing into the page will cause a force pushing it upward." This incorrect option continued to be a favorite among students and 10% chose option C after peer discussions.

Q24 is related to the direction of forces between two parallel wires carrying current in the same direction. The wires have different magnitudes of current flowing through them. The individual score for this item was 74% and the group score was 90%. It had an effect size of 0.39. This item had a construction rate of 89% and a co-construction rate of 0%. Only one group had an opportunity for co-construction in this item. Most students used Newton's third law to infer that the forces on both the wires will be the same. However, Table 2 shows that 21% of the students chose option D thinking that the force between the two wires will be repulsive. Some of these students appear to have made an analogy with the like charges repelling each other without taking the right-hand rule into consideration. For example, one upper-level student explained, "The force is repulsive because the charge flow is in the same direction."

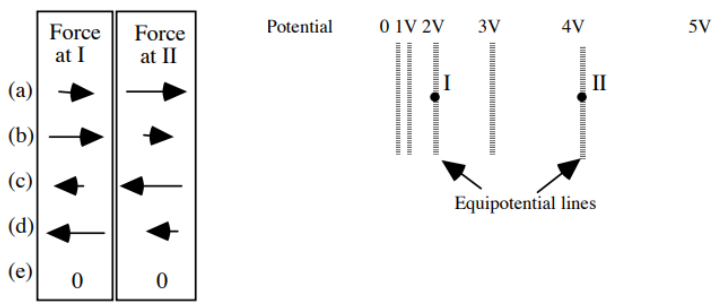

**Figure 9.** Figure provided with CSEM item 20

Q20 asks students to compare the electric force on a proton at two different points shown in figure 9. This item had an effect size of 0.38 with an individual score of 57% that increased to a group score of 75%. It had a construction rate of 70% and a co-construction rate of 50%. Table 2 shows that 14% of students chose option B and 19% chose option C incorrectly when answering the questions individually. Four groups had an opportunity to co-construct in Q20 but only two groups converged to the correct option D after peer interaction. The two groups who chose incorrect answers converged to incorrect option C with one student in each group providing the same incorrect option individually. It is evident that the concept of electric force when equipotential surfaces are shown diagrammatically remains a challenge for the students at all levels. Some students who chose option B explained their choices stating that the proton moves from a small to a high potential region. One student who chose option B explained, "Positive charges move from smaller to higher potential. The force will be greater when the equipotential lines are closer together." On the other hand, one of the students who chose option C explained, "Protons have a tendency to move from high potential to small potential, so the force will point left. The proton will feel a greater force at a higher potential." This shows that some students did not take the electric field into consideration correctly. Thus, most groups were able to converge on the correct direction for the forces but struggled with the determination of the magnitude of the electric forces.

Q16 investigates student understanding of the motion of an electron in a region in which information about the electric potential is provided only at one point (the question says that the potential is 10V there). This item also had a small effect size of 0.35 but it had an individual score of 83% which improved to 95% after group discussion. This item also did not have any opportunity for co-construction; however, there was a high construction rate of 86%. The most common incorrect answers were option B (the electron will move right (+x) since it is negatively charged, 7%) and option A (the electron will move left (-x) since it is negatively charged, 5%). Written explanations of the upper-level students who chose option B show that some of them thought that the electric potential increases in the positive x direction. One student explained, "I'm assuming the volts increases as you move in the positive x direction and opposites attract".

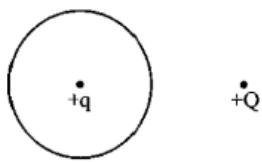

**Figure 10.** Figure provided with CSEM item 14

Q14 was the most difficult question among all the CSEM questions for the graduate students. This item is related to electrical shielding. Q14 has an isolated uncharged hollow conducting sphere and there is a charge +q inside the spherical shell (see figure 10). The question asks about the force experienced by the charges inside and outside the sphere. The average individual score was only 26%, which was the lowest individual score. The group score was 40% with an effect size of 0.30, construction rate of 70% and co-construction rate of 10%. This item is related to electrical shielding. 48% of the students considered Newton's third law and chose option A which says that the forces on the charges are equal and opposite to each other. The written explanations show that the upper-level undergraduate students thought that the sphere being neutral would not affect the forces so they will be equal. One of the students explained, "I don't believe the sphere has any effect because it is uncharged but Newton's laws say they will experience an equal and opposite force, and they are both positive so it would be away from each other". It is interesting that 50% of the groups chose the same option after discussion with their peers (see Table 2). Out of the 10 groups who had an opportunity for co-construction, only one group converged to the correct option D after group discussion. All the other groups gave an incorrect answer with 7 groups converging to the incorrect option A after discussion with their peers. This shows that graduate students struggle with the concept of electrical shielding and the instructors should use evidence-based approaches to help students develop a functional understanding of these concepts. Table 2 also shows that 12% of the students chose option B individually which shows they thought that neither charge will feel a force (e.g., thinking that the sphere will provide shielding such that neither feels a force). Moreover, Table 2 shows that 12% of the students chose option E individually which stated that both charges experience a net force but they are different from each other. These alternate conceptions were chosen by 5% of the groups after peer discussion. Written explanations from the upper-level students show that some students used Newton's third law, but they thought that the sphere may reduce the force on one of the charges. One student who chose option E explained, "'+q's charge will be distributed along the surface of the metal sphere (Faraday Shield) decreasing its charge density, +Q will feel less force. +q will instead still be subject to all of +Q but its own charge distributed to the sphere will cancel out".

**Small Effect Size**

Table 2 shows that there were 10 items with small effect size ($d < 0.3$) – Q1, 2, 4, 8, 9, 11, 12, 13, 28, 30. Analysis of items Q2, 11, 13, 28 and 30, which have an individual score lower than 85%, shows the characteristics of these items with small effect sizes. The small effect size of all the other items were due to the high individual scores so that there is less room for improvement during peer interaction.

Q11 evaluates student understanding of the change in the electric potential energy of a positive charge released from rest in a uniform electric field. It had a low individual score of 67% and a group score of 80%. The effect size was 0.28 with a construction rate of 75% and co-construction rate of 33%. Individually, 17% of the students chose incorrect option A (it will remain constant because the electric field is uniform) and 7% chose incorrect option C (it will increase because the charge will move in the direction of the electric field). Each of these incorrect options was chosen by 10% of the groups after discussion with their peers. Three groups had an opportunity to co-construct in Q11 but only one group gave the correct answer E after discussion with peers. One group consisted of two students who individually chose incorrect option A and stuck with the same option after their discussion. In the second group, one student individually selected incorrect option B while another chose incorrect option C, but they ended up choosing option C after discussing it. The written explanations show that upper-level students often chose option A because they incorrectly thought that the electric potential energy is proportional to the electric field. Those who chose option C appeared to incorrectly think that the electrical potential energy of the positive charge in a uniform electric field will increase because the charge will move in the direction of the field. This difficulty did not decrease after group discussion which shows that some students struggled with the concept of the motion of charges in a uniform electric field if the problem is posed in terms of change in electrical potential energy.

Q28 investigates the direction of the magnetic field at midpoint P midway on the vertical axis between two horizontal circular wire loops placed one above the other with the same vertical line passing through their centers, each carrying a current in counterclockwise direction. This item had an individual score of 74% and it went to 85% after peer interaction. It had an effect size of 0.26. It had an 86% construction rate and 0% co-construction rate. There were two groups with both students with incorrect individual answers, and they could not converge on the correct response as a group, which suggests that this could be a difficult concept for the students. The most dominant incorrect answer was option E (zero) provided by 26% of the students and it was reduced to 15% after group work. This shows that some students incorrectly thought that the contribution due to the magnetic field of the two wire loops get cancelled at point P. It appears that some students drew an analogy with two parallel straight wires carrying current in the same direction to conclude that the magnetic field would be zero at the center. But they did not realize that the magnetic field due to both loops points in the upward direction using the right-hand rule.

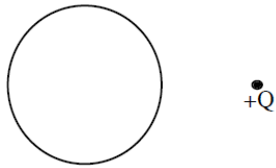

**Figure 11.** Figure provided with CSEM item 13

Q13 investigates student understanding of the direction of the electric field inside a hollow conducting sphere in the presence of an external positive point charge as shown in figure 11. This was the item with the third lowest effect size of 0.14. It had an average individual score of 74% which increased to 80% after peer discussion. This item had a 67% construction rate but 0% co-construction rate. Table 3 shows that there was only one group for which both students had incorrect answers, and they did not co-construct the correct answer. Individually, 17% of the students thought that the direction of the electric field is towards the left (option A) and 10% of the students thought it to be towards the right (option B). Table 2 shows that there was not much improvement after the peer interaction and 15% still chose option A and 5% chose B. Most of the students who chose incorrect answers did not realize the implications of the question asking about the electric field inside the conducting sphere in equilibrium. The electric field is zero within the conducting material in electrostatic equilibrium due to the shielding effect. Some upper-level students who provided written explanations only considered the induced surface charges to find the direction of the electric field. For example, one student who chose A explained, "all of the negative charges will accumulate on the right and the remaining positive charges will go to the left and the electric field will go from right to left" and another student who chose B explained, "Negative charge will distribute itself around the surface by where +Q is, so the positive charge will distribute to the left. Field lines will point from positive to negative, so this would be from left to right."

**Table 3.** Distribution of the number of groups whose responses on each question went from both incorrect individually to correct group response 00,1, both incorrect individually to incorrect group response 00,0, both correct individually to correct group response 11,1, one correct and one incorrect individually to incorrect group response 10,0, and one correct and one incorrect individually to correct group response 10,1 etc. (Number of groups = 20)

|     | 00,0 | 01,0/10,0 | 11,0 | 00,1 | 10,1/01,1 | 11,1 |
|-----|------|-----------|------|------|-----------|------|
| Q1  | 1    | 0         | 0    | 0    | 4         | 15   |
| Q2  | 1    | 2         | 0    | 0    | 3         | 14   |
| Q3  | 0    | 0         | 0    | 0    | 0         | 20   |
| Q4  | 0    | 0         | 0    | 0    | 2         | 18   |
| Q5  | 0    | 0         | 0    | 0    | 3         | 17   |
| Q6  | 0    | 0         | 0    | 0    | 4         | 16   |
| Q7  | 0    | 0         | 0    | 0    | 8         | 12   |
| Q8  | 0    | 1         | 0    | 0    | 4         | 15   |
| Q9  | 0    | 0         | 0    | 0    | 2         | 18   |
| Q10 | 0 | 0 | 0 | 0 | 6 | 14 |
| Q11 | 2 | 2 | 0 | 1 | 6 | 9 |
| Q12 | 0 | 0 | 0 | 0 | 2 | 18 |
| Q13 | 1 | 3 | 0 | 0 | 6 | 10 |
| Q14 | 9 | 3 | 0 | 1 | 7 | 0 |
| Q15 | 0 | 1 | 0 | 0 | 8 | 11 |
| Q16 | 0 | 1 | 0 | 0 | 6 | 13 |
| Q17 | 0 | 0 | 0 | 0 | 4 | 16 |
| Q18 | 2 | 0 | 0 | 0 | 9 | 9 |
| Q19 | 0 | 0 | 0 | 0 | 11 | 9 |
| Q20 | 2 | 3 | 0 | 2 | 7 | 6 |
| Q21 | 1 | 0 | 0 | 2 | 9 | 8 |
| Q22 | 1 | 2 | 0 | 0 | 10 | 7 |
| Q23 | 0 | 0 | 0 | 0 | 5 | 15 |
| Q24 | 1 | 1 | 0 | 0 | 8 | 10 |
| Q25 | 0 | 0 | 0 | 0 | 5 | 15 |
| Q26 | 0 | 0 | 0 | 0 | 3 | 17 |
| Q27 | 0 | 1 | 0 | 0 | 9 | 10 |
| Q28 | 2 | 1 | 0 | 0 | 6 | 11 |
| Q29 | 5 | 0 | 0 | 0 | 10 | 5 |
| Q30 | 1 | 2 | 0 | 0 | 4 | 13 |
| Q31 | 1 | 1 | 0 | 2 | 9 | 7 |
| Q32 | 5 | 1 | 0 | 0 | 10 | 4 |

Q30 investigates knowledge of Faraday's law using the concept of relative motion between a long straight wire (cylindrical shape) and a rectangular metal loop which moves with a speed v in different directions as shown in figure 12. The students were asked to determine those loops that have an induced current. This item had the second lowest effect size of 0.10 (the individual score for this item was 81% which increased to 85%). There was a construction rate of 67% but no co-construction was observed. Table 3 shows that there was only one group for which both individuals did not know the correct answer and they could not converge on the correct answer even after peer interaction. Table 2 shows that the most common incorrect choice was option D (all of the above) selected by 12%, which shows that many students knew that both I and II were correct in figure 12. Table 2 also shows that this answer choice decreased to 5% after peer interaction but option E (none of the above) increased from 5% to 10%, which is somewhat surprising. Thus, some students thought that there was no current induced in any of the loops after discussions with their peers. This shift from the individual to group choice suggests that group interaction can also have a negative impact if students are not comfortable with the concepts and the difficulty level of the problem is beyond a particular group's ZPF [5, 7].

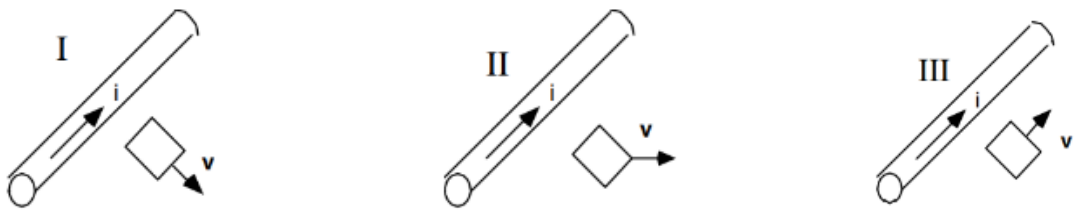
**Figure 12.** Figure provided with CSEM item 30

Q2 investigates students' knowledge of the distribution of charges on insulators. It asks students what happens to a small amount of negative charge placed at a point P on the outside of an insulating hollow sphere. It was the item with the lowest effect size of 0.05. There was a 60% construction rate but a 0% co-construction rate. Table 3 in the Appendix shows that there was only one group with both incorrect individual choices, and they were not able to converge on the correct answer as a group. On Q2, 83% of the students provided the correct response individually (option A), i.e., all the excess charge remains right around P whereas 5% chose option D (most of the charge is still at point P but some of the charge will spread over the sphere) and 7% chose option E (there will be no excess charge left). After the peer interaction, 85% of the groups chose the correct answer and 15% chose option D. Written explanations from students who chose option D suggest that they considered the insulating nature of the sphere, but it was unclear from those explanations as to why some charges would spread out over the sphere. We note that students performed better on the preceding item Q1 in which the only difference was that the sphere was conducting. On Q1, the score increased from 86% to 95% after the group work.

**RQ5. How does the group performance of the graduate students compare to the group performance of the introductory students on the CSEM?**

Comparison of the group performance of graduate students in this study with the introductory students in prior study [43] shows that collaboration was productive in both cases. For example, the rates of construction were 88% and 81%, respectively, for the graduate and introductory students. The rates of co-construction were 19% and 29%, respectively, for the graduate and introductory students. Moreover, prior research shows that out of 32 questions, there were 8 questions on the CSEM survey with a group score less than 65% for introductory physics students [43]. We compared the group scores on these items (Q14, 18, 20, 21, 22, 27, 31, 32) between introductory and graduate students to investigate if the concepts got easier or remained challenging for the graduate students and how the group scores for the two groups were impacted by peer interaction. Table 4 in the Appendix shows that the group score for Q14 was 48% for introductory students and 40% for graduate students. It was the lowest group score for the graduate students which shows that the concept of electrical shielding is the most challenging at both levels. This was the only item in which less than 70% of graduate student groups provided the correct answer after working with their peers. The rates of construction and co-construction were also similar at both levels. In particular, on Q14, the construction rate was 63% and 70% in introductory and graduate level, respectively. The rate of co-construction (10%-12%) was almost the same for students at both levels as shown in Table 4. Q20 and Q32 had group scores of 40% and 43%, respectively, for the introductory student groups (see Table 4). The graduate student groups performed comparatively better on these two questions with group scores of 75% and 70%. This indicates that the concepts related to the electric force on charged particles (Q20) and Faraday's law (Q32) became easier to some extent for the graduate students. Both the rates of construction and co-construction for Q20 were roughly 20 percentage points higher for graduate students compared to introductory students. However, Table 4 shows that for Q32, the rate of construction was roughly 30 percentage points higher for graduate students compared to introductory students, but the rate of co-construction was 26% for introductory students and 0% for graduate students. In fact, Q18, 21, 22, 27 and 31 were among the items on which the graduate students showed an excellent group performance such that all the items had a group score ≥ 85%. It is interesting to note that Q21 and Q31 were two of the items with low group scores of 55% and 52% for introductory students. However, these items were in the range of high effect size for graduate students with a group score of 95% and 90%, respectively. This was due to the fact that on Q21, the rate of co-construction was 10% for introductory students but 67% for graduate students, and on Q31, it was 24% for introductory students but 67% for graduate students. On Q18, 22 and 27, the group scores of the graduate students were approximately 30% higher compared to the introductory level. This was mostly due to the increased rate of construction for all three items for graduate students compared to the introductory students. Table 4 shows that the rates of co-construction for Q18 and Q22 are 0% each for the graduate students (two groups for Q18 and one group for Q22-see Table 3). Table 4 shows that Q27 had no opportunity for co-construction at the graduate level as there were no groups for which both students provided an incorrect answer individually. These items covered the concepts related to electric field, and magnetic field and force. Table 4 shows that even though introductory students struggled with these concepts, first year graduate students appeared to have a robust knowledge of these concepts.

**Table 4.** Comparison of group scores in percentage for introductory students [43] and graduate students along with rates of construction and co-construction for each survey item. "-" refers to no opportunity for co-construction i.e., there were no groups in which all students gave an incorrect answer individually.

| | Introductory | | | Graduate | | |
|---|---|---|---|---|---|---|
| Q | Group Score | Construction | Co-Construction | Group Score | Construction | Co-Construction |
| 1 | 99 | 100 | 0 | 95 | 100 | 0 |
| 2 | 87 | 71 | 11 | 85 | 60 | 0 |
| 3 | 95 | 92 | 0 | 100 | - | - |
| 4 | 80 | 73 | 33 | 100 | 100 | - |
| 5 | 84 | 86 | 25 | 100 | 100 | - |
| 6 | 93 | 83 | 25 | 100 | 100 | - |
| 7 | 77 | 83 | 36 | 100 | 100 | - |
| 8 | 95 | 87 | 0 | 95 | 80 | - |
| 9 | 88 | 78 | 100 | 100 | 100 | - |
| 10 | 82 | 88 | 39 | 100 | 100 | - |
| 11 | 65 | 80 | 29 | 80 | 75 | 34 |
| 12 | 93 | 85 | 50 | 100 | 100 | - |
| 13 | 76 | 83 | 16 | 80 | 67 | 0 |
| 14 | 48 | 63 | 12 | 40 | 70 | 10 |
| 15 | 72 | 81 | 35 | 95 | 89 | - |
| 16 | 77 | 76 | 45 | 95 | 86 | - |
| 17 | 70 | 69 | 16 | 100 | 100 | - |
| 18 | 57 | 53 | 34 | 90 | 100 | 0 |
| 19 | 77 | 80 | 55 | 100 | 100 | - |
| 20 | 40 | 51 | 30 | 75 | 70 | 50 |
| 21 | 55 | 85 | 10 | 95 | 100 | 67 |
| 22 | 54 | 55 | 28 | 85 | 83 | 0 |
| 23 | 92 | 95 | 56 | 100 | 100 | - |
| 24 | 66 | 83 | 23 | 90 | 89 | 0 |
| 25 | 84 | 83 | 27 | 100 | 100 | - |
| 26 | 97 | 100 | 63 | 100 | 100 | - |
| 27 | 57 | 72 | 10 | 95 | 90 | - |
| 28 | 80 | 85 | 40 | 85 | 86 | 0 |
| 29 | 71 | 78 | 44 | 75 | 100 | 0 |
| 30 | 65 | 61 | 29 | 85 | 67 | 0 |
| 31 | 52 | 91 | 24 | 90 | 90 | 67 |
| 32 | 43 | 62 | 26 | 70 | 91 | 0 |
| | Average | 78 | 29 | | 88 | 19 |

## 5. SUMMARY AND INSTRUCTIONAL IMPLICATION

This study shows that the graduate students who worked individually on the CSEM survey improved their performance as a group without any guidance from the instructor. All the questions had a group score ≥ 70% except for Q14 which involved the most challenging concept at the graduate level. Most of these items were in the range of small and medium effect size with an average construction rate of 88%. This suggests that the groups in which one student was correct, and the other was incorrect, converged to a correct answer as a group 88% of the time. Most CSEM questions were in the zone of proximal facilitation as evidenced by the high rate of construction. However, the fact that construction rate is not 100% points to the fact that peer discussion without guidance from the instructor can sometimes lead to students converging on incorrect responses as a group. On average, the rate of co-construction was observed to be 19%, with co-construction occurring in eight instances across five items (even though there was opportunity for co-construction in 15 items). Our findings indicate that in 75% of these eight instances, both members of a group initially had different incorrect answers (i.e., in 25% of these instances, both members had the same incorrect answer) but were able to converge to the correct response after engaging in discussion with their peers.

Moreover, the group score of the graduate students was compared with the group score of the introductory students [43] which showed that there is consistency in the performance of the students at both levels. The concepts that were challenging to introductory students were the ones on which the graduate students also struggled in the individual administration of the survey. But the scores were generally higher for the graduate students.

There were a few concepts, e.g., related to the shielding effect in conductors and Faraday's Law and Lenz's law which were very challenging for graduate students and the improvement after peer interaction was not as large as one may have hoped. In particular, these questions were out of the zone of proximal facilitation for many of the groups and students did not benefit from group interaction. For example, Table 3 shows that on Q14 about electrical shielding, 10 groups of graduate students had both group members with incorrect response individually but only one of those groups was able to converge on the correct answer after discussion with peers. Similarly, Table 3 shows that on items 29 and 32 about Faraday's law and Lenz's law, 5 groups of graduate students had both group members with incorrect response individually for each question but none of those groups was able to converge on the correct answer after discussion with peers. The fact that the graduate students could not converge on the correct answer to those CSEM questions even after discussing them with their peers suggests that these questions deserve special attention from physics instructors. It appears that like introductory students, the graduate students have strong alternative conceptions that prevent them from getting to the correct conclusion while answering these questions even after peer interaction. Therefore, it is important for the instructors to focus on the evidence-based approaches to help students learn about these concepts involving conductors/insulators and Faraday's law and Lenz's law. There is a need for developing and implementing research-based curricula and pedagogies, e.g., tutorials, and think-pair-share questions to help students develop a functional understanding of the underlying concepts so that there is little room for alternative conceptions. These types of research-based curricula can be integrated with group work to create a cognitive conflict that can lead to productive discussions among the students during the group work. Overall, our findings suggest that physics instructors should incorporate group interactions both inside and outside of the classroom even without the instructor involvement so that students at all levels can learn from each other and develop a functional understanding of the underlying concepts.

A key limitation of our study is the lack of insight into individual participation within groups. Even though students knew each other (being in the same cohort), chose their partners and were asked to engage deeply in group discussion, it's possible that at least in some groups, some students dominated the discussion, while others simply agreed with their peers. Future studies should investigate how individual students' contributions vary within groups to get a clearer picture of group dynamics and learning outcomes. Also, while this investigation aimed to evaluate the differences between individual performance and group performance, it would be interesting to investigate whether the improvement in group performance were due to peer collaboration or because students had more time to reflect on the questions individually. Future research should compare a group of graduate students working together with a control group who retake the survey individually to understand how similar or different are their improvements. Additionally, our focus was on first-year graduate students who likely have a strong sense of community as they have similar experiences and take multiple courses together during the semester. This shared academic journey along with the fact that they were allowed to choose their partners may have positively influenced their peer collaborations. It would be useful to investigate how peer collaborations may be affected if students working in groups do not have the same sense of community.

**Ethical statement**

This research was carried out in accordance with the principles outlined in the University of Pittsburgh Institutional Review Board (IRB) ethical policy, the Declaration of Helsinki, and local statutory requirements. Informed consent was obtained from all interviewed students who participated in this investigation and for the results to be published.